# FASL-Seg: Anatomy and Tool Segmentation of Surgical Scenes

Muraam Abdel-Ghani[a,*], Mahmoud Ali[b,1], Mohamed Ali[b,1], Fatmaelzahraa Ahmed[a], Mohamed Arsalan[b],
Abdulaziz Al-Ali[b] and Shidin Balakrishnan[a,**]

[a]Department of Surgery, Hamad Medical Cooperation
[b]College of Engineering, Qatar University
ORCID (Muraam Abdel-Ghani): https://orcid.org/0000-0001-5876-6648, ORCID (Mohamed Ali):
https://orcid.org/0009-0003-7242-585X, ORCID (Shidin Balakrishnan): https://orcid.org/0000-0001-6361-4980

**Abstract.** The growing popularity of robotic minimally invasive surgeries has made deep learning–based surgical training a key area of research. A thorough understanding of the surgical scene components is crucial, which semantic segmentation models can help achieve. However, most existing work focuses on surgical tools and overlooks anatomical objects. Additionally, current state-of-the-art (SOTA) models struggle to balance capturing high-level contextual features and low-level edge features. We propose a Feature-Adaptive Spatial Localization model (FASL-Seg), designed to capture features at multiple levels of detail through two distinct processing streams, namely a Low-Level Feature Projection (LLFP) and a High-Level Feature Projection (HLFP) stream, for varying feature resolutions - enabling precise segmentation of anatomy and surgical instruments. We evaluated FASL-Seg on surgical segmentation benchmark datasets EndoVis18 and EndoVis17 on three use cases. The FASL-Seg model achieves a mean Intersection over Union (mIoU) of 72.71% on parts and anatomy segmentation in EndoVis18, improving on SOTA by 5%. It further achieves a mIoU of 85.61% and 72.78% in EndoVis18 and EndoVis17 tool type segmentation, respectively, outperforming SOTA overall performance, with comparable per-class SOTA results in both datasets and consistent performance in various classes for anatomy and instruments, demonstrating the effectiveness of distinct processing streams for varying feature resolutions.

## 1 Introduction

Robot-assisted surgeries are becoming more widely adopted due to their numerous benefits, such as fewer post-operative complications and faster recovery times [12]. However, the complexity of these robotic systems can challenge novice surgeons [14], requiring extensive training. Advances in artificial intelligence (AI) have facilitated a deeper understanding of the surgical environment by accurately identifying instruments and anatomical features. Precise localization is crucial for surgical training, allowing for constructive feedback and objective skill assessments [2]. To support this, semantic segmentation of the surgical scene [11] enables pixel-level labeling for locating tools and anatomy. However, challenges arise in feature extraction due to differences in the representation of various objects. Low-level features encoding tool edges and small objects are extracted from earlier blocks of the encoder, while contextual high-level features encoding anatomy and larger tools are extracted in the later stages [15]. Achieving effective detail preservation and enhanced anatomical representation necessitates distinct processing of these multiscale features.

We propose a Feature-Adaptive Spatial Localization (FASL-Seg) model, a multiscale segmentation architecture that adapts feature processing using separate streams for low and high-level features. This approach maintaines information integrity, ensuring fine-grained and precise localization of surgical tools and anatomy. The architecture is built on a transformer backbone, specifically the SegFormer [24] model. The Low-Level Feature Projection (LLFP) stream enhances edge information extracted from early layers, while the High-Level Feature Projection (HLFP) stream processes contextual features encoded in later model layers.

- We propose a novel HLFP and LLFP-powered multiscale segmentation architecture that captures variational features with enhanced contextual understanding,
- Our proposed method is the combination of low-level features and high-level features that include edge information from initial layers of the network, which results in better segmentation performance,
- Our proposed architecture aggregates HLFP and LLFP in combination with a shallow decoder for both tools and anatomy segmentation
- The code and trained models are available at [1]

## 2 Related Work

Various studies proposed modules and architectures designed to improve the precision and effectiveness of surgical segmentation outputs. With the emergence of transformer models, many researchers explored integrating these advanced models into their architectures. MedT was proposed by [21] with two segmentation branches, a global and a local branch incorporating transformer blocks. The global branch includes only two blocks compared to five in the local branch. The model also uses a gated axial attention mechanism in the transformer implementation. Finally, the two branch outputs are merged and upsampled through the decoder. The authors of [7] proposed a hybrid transformer-unet architecture called TransUNet, with two sequential encoders, and the output of the transformer encoder is subsequently merged with the U-Net encoder block outputs. Although it performed well in small object segmentation and contextual feature extraction, it was found to struggle with thin objects such as needle and thread when tested for surgical scene segmentation [16].

* Corresponding Author. Email: muraam.abdelghani@outlook.com
** Corresponding Author. Email: sbalakrishnan1@hamad.qa
[1] Equal contribution.

In [23], the authors also proposed a hybrid Vision Transformer (ViT) and convolutional neural network (CNN) architecture. They use two encoder paths, one with transformer blocks and the other with CNN blocks. An adapter module was proposed to apply cross-communication between the feature maps extracted from the two paths. The cross-communication insights are merged with the transformer and CNN encoder outputs and processed to produce the final model output. The model showed better performance compared to SOTA. However, it still struggled with some surgical tools such as the clip applier, grasping retractor and ultrasound probe. The authors of [5] proposed a Masked-Attention Transformer model (MATIS) based on a Mask2Former backbone. It utilizes a pixel decoder and a transformer decoder, and introduces a temporal consistency module utilizing temporal information from consecutive frames to improve the classification of surgical tools. While these components improved SOTA results, using multiple backbones and decoders increases model complexity. Unlike these previous methods, our model uses only one encoder based on a SegFormer backbone, and one decoder, and incorporates the distinct feature processing streams for each encoder output to merge them together, rather than processing and incorporating them into the feature map gradually. By this, our work aims to retain maximum information from various feature map resolutions.

Notably, most previous methods focus on instrument segmentation rather than full surgical scene segmentation. Additionally, work that included anatomy and tool segmentation rarely reported per-class performance metrics that reveal performance variations between these classes. One paper [16] proposed a Segment Anything Model (SAM)-based approach with a text affinity module to transform text prompt embeddings for medical terms. They reported results for transformer-based models, namely MedT and TransUNet, on most classes of the EndoVis18 dataset, which includes instruments and anatomy labels, but combined the instrument shaft, wrist, and clasper classes into one "robotic instrument" class. Another paper that reported per-class results for EndoVis18 is [11], which concentrated on semantic surgical video segmentation using inter-frame relations, and was built upon a SwinTransformer backbone. However, it only reports per-class results for 9 classes of the dataset. In our work, we address this gap in the literature by reporting per-class metrics for all tools, tool parts, and anatomical structure classes present in the EndoVis18 parts and anatomy dataset.

## 3 Method

### 3.1 Design Motivation

The output of the model encoder blocks typically follow a pattern where the feature map resolution decreases while the channel-encoded data increases. This results in a transition from capturing fine edge details to broader contextual information, as highlighted in [15]. Figure 1 presents a heatmap illustrating the activated outputs from the first two encoder blocks of the SegFormer [24] model. The first encoder block focuses on detailed edge features and precisely highlights tool-tips, while the second block encodes essential low-level features of smaller objects, such as the instrument claspers and wrists, with limited contextual information. Conversely, the output of the third and fourth encoder blocks, depicted in Figure 2, showcase smaller feature map sizes that capture global and context-related characteristics. This comparison clearly indicates that to enhance the model's ability to segment entire surgical scenes effectively, it is necessary to process low-level and high-level features distinctly, ensuring that the segmentation model output retains multiscale feature representations. This approach could be achieved through the proposed feature processing streams that will be discussed in the next section.

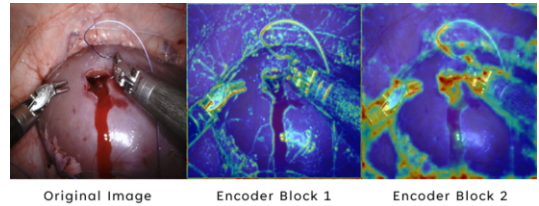

**Figure 1.**　Output activation heatmaps for First and Second SegFormer Encoder Block output activations

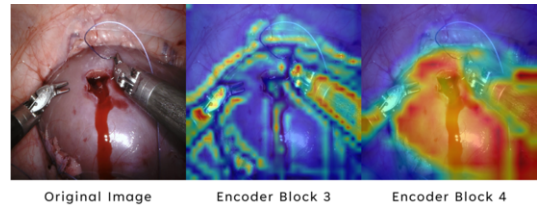

**Figure 2.**　Output activation heatmaps for Third and Fourth SegFormer Encoder Block output activations

### 3.2 Overall Architecture

In this section, we present a detailed description of the proposed architecture for FASL-Seg. The model encoder is built on the SegFormer backbone [24], which is a transformer model that features hierarchical encoder blocks with different spatial dimensions. With its attention mechanisms and overlapping nature of patch embeddings, the transformer encoder provides powerful feature extraction for semantic scene understanding, particularly in medical applications[7]. This is further supported by the fact the SegFormer model eliminates the positional encoding of patches. This feature makes our segmentation model more robust against variations in input frame resolutions, which can arise from different video capture technologies used by various surgical robotic platforms. To leverage the strengths of SegFormer, we integrate it into our model. The encoder outputs are processed through two separate streams: a low-level feature projection (LLFP) and high-level feature projection (HLFP) stream, each applying distinct operations based on the resolution of the feature maps. This approach not only preserves local features, but also captures contextual information effectively. Ultimately, the refined multi-scale features are processed in the shallow model decoder to produce the final segmentation output.

### 3.3 LLFP Stream

To preserve the details of the first and second encoder feature maps, we propose a Low-Level Feature Projection (LLFP) stream to process local information. Given a feature map output of the $i^{th}$ encoder block, $F_i$. Firstly, the feature map is passed through a Point-wise Convolution (PWConv) layer, followed by a Batch Normalization and Leaky ReLU; hereafter, we shall refer to this combination as a ConvBlock. The PWConv layer allows spatial dimensions to be maintained while refining the feature representations. This is aided by using a small kernel size of 1, which prevents excessive smoothing of the fine details. The block's output is represented by:

$$ConvBlock(F_i) = LR(BN(Conv_1(F_i))) \quad (1)$$

where $Conv_1$ is the PWConv layer, $BN$ stands for Batch Normalization, and $LR$ stands for Leaky ReLU activation. The output of this block is then passed to a Multi-Head Self Attention (MHSA) block described in [22], represented by the following equations, where $Q$, $K$, and $V$ are query, key and value vectors, $W$ represents the output weight matrix, and $d_k$ the key dimension scaling factor:

$$MHA(Q, K, V) = Concat(head_1, ..., head_h)W^O \quad (2)$$

$$head_i = Attention(QW_i^Q, KW_i^K, VW_i^V) \quad (3)$$

$$Attention(Q, K, V) = softmax(QK^T/\sqrt{d_k})V \quad (4)$$

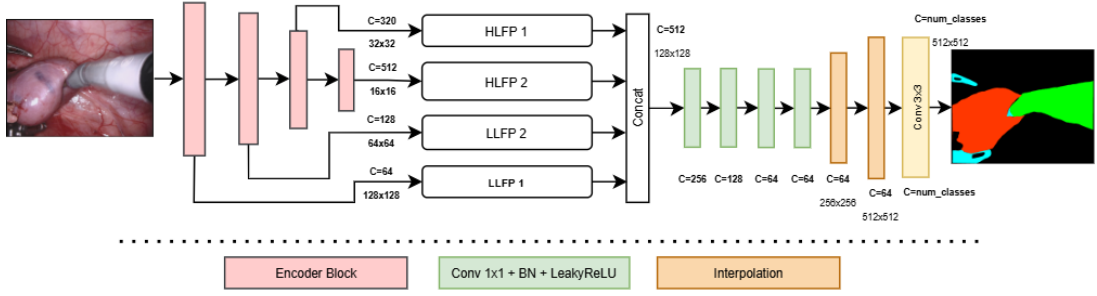

**Figure 3.** Overview of Proposed Architecture for FASL-Seg

The feature map is passed as the query, key, and value. With several heads, it is possible to enhance the details represented in the feature map, as multiple representations of the feature map can be learned concurrently. Furthermore, with higher resolution feature maps, local and global dependencies can be captured by the MHSA block, and irrelevant noise can be removed. The output of this block can thus be represented as:

$$\hat{F}_i = MHSA(ConvBlock(F_i))) \quad (5)$$

The ConvBlock and MHSA constitute the main components of the LLFP stream. To prepare the feature maps for channel-wise concatenation, interpolation is added to the LLFP stream where necessary for feature map enlargement, specifically for the second encoder block output. Multiple interpolations are used to prevent feature abstraction from enlarging the feature map to a large scale at once, each enlarging the feature map size by a factor of 2. This is represented by equation 6, hereafter called the Up Chain, and the final LLFP stream output is represented by equation 7:

$$UpChain_N(F_i) = Up_N(...Up_1(F_i)) \quad (6)$$

$$\hat{F}_i = UpChain_N(MHSA(ConvBlock(F_i))) \quad (7)$$

### 3.4 HLFP Stream

Given a feature map, $F_i$, output from the $i^{th}$ encoder block. Like the LLFP, the feature map is passed to a chain of Conv blocks, preserving the extracted contextual features, while enabling compression of channel-wise features into fewer channels. This block's output can thus be represented by equation 8, where $ConvBlock$ is the block defined by equation 1.

$$ConvChain_N(F_i) = ConvBlock_N(...ConvBlock_1(F_i)) \quad (8)$$

Unlike in LLFP, where attention is needed to minimize noise and enhance the detailed features, the HLFP does not use MHSA, as the extracted features are already high-level and capture the global context with little noise. The introduction of attention would thus compromise essential features present in this stage. As a result, the output of the Conv block is passed directly to enlargement with interpolation. The final shape of the HLFP stream output is presented in equation 9.

$$\hat{F}_i = UpChain_N(ConvChain_N(F_i)) \quad (9)$$

### 3.5 Final Architecture of Model

Based on the output of each encoder block, the appropriate configuration of LLFP or HLFP stream components was used. Since the model is based on a SegFormer encoder, there are four encoder blocks. For the first and second encoder block outputs, LLFP was used. For the first LLFP stream, the interpolation was set as N=0, since the feature map size of 128×128 does not require further enlargement (eq. 10). On the other hand, the second LLFP required interpolation with N=1 as the feature map size was 64×64 (eq. 11). HLFP was used for the output of the third and fourth encoder blocks. For the output of the third block, the choice of N for the Conv Chain in the stream is 1 and for the Up Chain is 2 (eq. 12). For the output of the fourth encoder block, the N for ConvChain is 2 and for Up Chain is 3 (eq. 13). These are chosen based on the required conversions from the initial encoder output size to the final output size that matches that of the LLFP streams outputs, specifically 128×128 with 128 channels. Equation 14 presents the fusion of the processed multiscale features from the four streams, producing the final enhanced merged feature map, $\hat{F}_{EM}$. A simplified representation of the final LLFP and HLFP streams used in the model architecture is depicted in Figure 4.

$$\hat{F}_1 = MHSA(ConvBlock(F_1)) \quad (10)$$

$$\hat{F}_2 = UpChain_1(MHSA(ConvBlock(F_2))) \quad (11)$$

$$\hat{F}_3 = UpChain_2(ConvChain_1(F_3)) \quad (12)$$

$$\hat{F}_4 = UpChain_3(ConvChain_2(F_4)) \quad (13)$$

$$\hat{F}_{EM} = Concat(\hat{F}_1, \hat{F}_2, \hat{F}_3, \hat{F}_4) \quad (14)$$

The shallow model decoder consists of four ConvBlocks to enable weighted feature selection and channel compression of the enhanced merged feature map. This ensures preservation of spatial details and edge information. The feature map size is enlarged using bilinear interpolation, and the output is passed to a final Laplacian convolution layer with a channel size corresponding to the number of classes, forming the final segmentation output. Figure 3 presents the proposed FASL-Seg architecture.

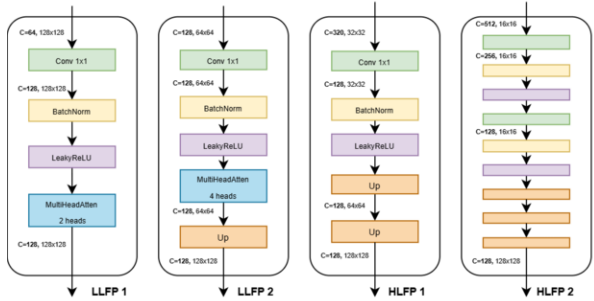

**Figure 4.** LLFP and HLFP streams in proposed architecture

## 4 Datasets and Experimental Setup

### 4.1 Datasets

To evaluate the efficacy of the proposed model architecture, the model was trained on two main datasets, namely EndoVis17 [3] and EndoVis18 [4] Challenge Datasets. Both datasets consist of frames from nephrectomy procedures completed using the da Vinci Xi minimally invasive surgical platform. EndoVis17 consists of 10 procedures with 300 frames each at dimensions 1280×1024. The official testing set consists of sequences 9 and 10, with a 25% split from each of the sequences 1 through 8. The training set consists of the remaining 75% split from the respective 8 sequences, where each contains 225 frames after splitting. The segmentation maps provide labels for surgical instruments. For the EndoVis18 dataset, the training set consists of 15 procedural sequences with 149 frames each, and the testing set consists of 4 sequences each with 249 frames, also at 1280×1024 pixels. The original challenge consisted of surgical instrument parts and anatomy. However, using the official training sequences, an additional set of annotations was published for the surgical tools specifically. For this version of the dataset, the sequences were split into training and testing following the approach in [9], with sequences 2, 5, 9 and 15 for testing and the rest for training. The train set was randomly split into 80% training and 20% validation sets. The fixed seed was used for all the experiments to ensure a consistent random split every time.

### 4.2 Experimental Setup

All models and baselines were trained in a Linux environment utilizing an A6000 GPU with 48 GB GPU RAM, 8 CPUs and 45 GB of RAM. Models were trained for 100 epochs, with a batch size of 4 due to memory constraints, and using an Adam optimizer with a fixed learning rate of 1E-5 for convergence stability, with no weight-decay. For the loss function, a combination of Tversky ($\mathcal{L}_{tversky}$) [18] and Cross Entropy loss ($\mathcal{L}_{CE}$) [13] was used, following the equation below:

$$\mathcal{L}_{tversky} = \frac{TP}{TP + \alpha FP + \beta FN} \quad (15)$$

$$\mathcal{L}_{total} = \alpha \mathcal{L}_{tversky} + (1-\alpha)\mathcal{L}_{CE} \quad (16)$$

In our experiments, for the tversky loss, the chosen $\alpha$ value is 0.7 and $\beta$ is 0.3, penalizing false predictions to prevent oversegmenting. For equation 16, the chosen $\alpha$ is 0.5, allowing equal contribution from both losses to the final model loss. For measuring the model performance, the reported metrics were mean Intersection over Union (mIoU) and Dice Similarity Coefficient (Dice), following the equations 17 [9] and 18 respectively. The value of $\epsilon$ is used to prevent division by zero, and is chosen to be 1E-7 for both equations. Only random resized cropping, horizontal flipping, and vertical flipping were applied for data augmentation, to replicate realistic scenarios where the camera might be flipped during operation or when structures are not completely visible. All images were resized to 512 by 512 for model input. However, results are based on resizing predictions back to their original dimensions to preserve details created by medical experts. This approach ensures an accurate indication of the model's performance, especially for small or thin objects.

$$\text{mIoU} = \frac{1}{N_c} \sum_{c=1}^{N_c} \frac{\text{Intersection}_c + \epsilon}{\text{Union}_c + \epsilon} \quad (17)$$

$$\text{Dice Mean} = \frac{1}{N_c} \sum_{c=1}^{N_c} \frac{2 \cdot \text{Intersection}_c + \epsilon}{\text{Union}_c + \epsilon} \quad (18)$$

## 5 Results

The performance of FASL-Seg was evaluated in the three usecases based on the datasets described in Section 4.1. For EndoVis18 parts and anatomy segmentation, little literature reported the per-class performance for all the original dataset classes. We compare the performance of FASL-Seg to reported benchmarks for transformer-based models trained on this task, namely TransUNet and MedT [16], and SOTA architectures including UNet[17], DeepLabV3 [8], MaskR-CNN [10] and the SegFormer-b5 variant trained following the same training hyperparameters presented in Section 4.2, with the goal of obtaining per-class results for the instrument shaft, wrist and clasper independently. For EndoVis18 tool segmentation, we benchmark against transformer-based models models proposed for surgical tool segmentation including Swin-Transformer based MATIS [5], TraSeTR [25], S3Net [6], MSLRGR [19] and ViTxCNN [23] as reported in [23], in addition to training and evaluating UNet and SegFormer-b5. Similarly, for EndoVis17, the benchmark models include UNet and transformer-based models for surgical segmentation MATIS, TraSeTR, S3Net, ViTxCNN and SegFormer-b5.

The results for EndoVis18 parts and anatomy segmentation in Table 1 and 2 demonstrate that FASL-Seg outperforms the benchmark models, exceeding the highest performing model, MedT, by 8% in mIoU and 9% in Dice. It also achieves comparable performance to SOTA on the rest of the classes. While some SOTA models surpass FASL-Seg in specific classes, it maintains the highest average per-class mIoU and Dice, indicating a more consistent performance. For EndoVis18 tool segmentation, the FASL-Seg model outperforms SOTA in overall metrics and classes *Large Needle Driver* and *Monopolar Curved Scissors*, achieving the best average per-class metrics. Likewise, for EndoVis17 tool segmentation, it surpasses SOTA for overall and several specific classes. This observation showcases the model's reliability across different feature representations.

The error analysis of the metrics per class for the combined anatomy and parts use case, shown in Figure 6 revealed high performance in classes such as the instrument shaft and ultrasound probe, while lower performance in some classes such as the instrument clasper and covered kidney, which can be addressed by in-troducing additional augmentation steps such as hue and contrast augmentation. To better visualize the different model inferences, we analyze some testing frame inference examples. Figure 5 presents visual comparison of the segmentation performance of FASL-Seg against SOTA models. The overlay images reveal that SegFormer and FASL-Seg achieved more precise anatomical segmentation compared to CNN-based models. UNet produced good segmentation for larger tool details such as enlarged instrument parts, but struggled with smaller tools/tips such as the upper tool claspers in the first frame, a trend also seen in MaskRCNN and DeepLabV3. Although SegFormer produced higher precision for anatomy, it struggled with the precise tool segmentation. In the first frame, SegFormer missed some tool tips such as the clasper on the right in the second frame, and oversegmented some areas such as the holes of the instrument clasper on the left of the third frame. FASL-Seg successfully captured all instrument tips, though it missed some of the covered kidney and kidney parenchyma areas. This points to potential improvements through additional augmentation techniques. Overall, FASL-Seg delivered higher precision segmentation maps across most classes, indicating the efficacy of the model in the segmentation of both anatomical classes and instrument parts.

**Table 1.** Mean IoU Results EndoVis18 Parts and Anatomy Segmentation. Label Key: BT: Background Tissue ISh:Instrument Shaft, IC: Instrument Clasper, IW: Instrument Wrist, KP:Kidney Parenchyma, CK: Covered Kidney, SmInst: Small Intestine, SI: Suction Instrument, UP: Ultrasound Probe

| Model | mIoU | BT | ISh | IC | IW | KP | CK | Thread | Clamps | Needle | SI | SmInt | UP | Avg |
|---|---|---|---|---|---|---|---|---|---|---|---|---|---|---|
| U-Net | 0.53 | 0.65 | 0.72 | 0.37 | 0.45 | 0.43 | 0.07 | 0.38 | 0.76 | 0.91 | 0.70 | 0.18 | 0.72 | 0.53 |
| Mask-RCNN | 0.37 | 0.68 | 0.82 | 0.40 | 0.56 | 0.64 | 0.18 | 0.03 | 0.45 | 0.00 | 0.00 | 0.43 | 0.22 | 0.37 |
| DeepLabV3 | 0.35 | **0.87** | 0.49 | **0.62** | 0.72 | 0.26 | 0.19 | 0.41 | 0.00 | 0.00 | 0.36 | 0.35 | 0.00 | 0.36 |
| SegFormer | 0.57 | 0.48 | 0.15 | 0.12 | 0.18 | 0.05 | **0.52** | 0.90 | 0.93 | 0.91 | **1.00** | 0.77 | 0.85 | 0.57 |
| TransUNet [16] | 0.48 | 0.59 | | 0.60 | | 0.32 | 0.33 | 0.01 | **1.00** | 0.04 | 0.64 | 0.63 | 0.61 | 0.48 |
| MedT [16] | 0.65 | 0.40 | | 0.54 | | 0.16 | 0.44 | 0.82 | 0.96 | 0.90 | 0.61 | 0.78 | 0.84 | 0.65 |
| FASL-Seg(Ours) | **0.73** | 0.79 | **0.85** | 0.53 | 0.65 | **0.66** | 0.29 | 0.69 | 0.90 | **0.91** | 0.85 | 0.77 | 0.84 | **0.73** |

**Table 2.** Dice Results for EndoVis18 Parts and Anatomy Segmentation. Label Key: BT: Background Tissue ISh:Instrument Shaft, IC: Instrument Clasper, IW: Instrument Wrist, KP:Kidney Parenchyma, CK: Covered Kidney, SmInst: Small Intestine, SI: Suction Instrument, UP: Ultrasound Probe

| Model | Dice | BT | ISh | IC | IW | KP | CK | Thread | Clamps | Needle | SI | SmInt | UP | Avg |
|---|---|---|---|---|---|---|---|---|---|---|---|---|---|---|
| U-Net | 0.58 | 0.78 | 0.78 | 0.50 | 0.55 | 0.54 | 0.11 | 0.38 | 0.77 | 0.91 | 0.70 | 0.20 | 0.73 | 0.58 |
| Mask-RCNN | 0.48 | 0.81 | **0.90** | 0.57 | 0.72 | **0.78** | 0.31 | 0.07 | 0.62 | 0.0 | 0.0 | 0.60 | 0.36 | 0.48 |
| DeepLabV3 | 0.46 | **0.93** | 0.66 | **0.76** | **0.84** | 0.42 | 0.31 | 0.58 | 0.00 | 0.00 | 0.53 | 0.52 | 0.00 | 0.46 |
| SegFormer | 0.58 | 0.64 | 0.12 | 0.12 | 0.18 | 0.05 | **0.52** | 0.90 | 0.93 | 0.91 | **1.00** | 0.77 | 0.85 | 0.58 |
| TransUNet [16] | 0.52 | 0.73 | | 0.70 | | 0.49 | 0.33 | 0.00 | **1.00** | 0.04 | 0.66 | 0.63 | 0.62 | 0.52 |
| MedT [16] | 0.68 | 0.56 | | 0.66 | | 0.26 | 0.44 | 0.82 | 0.96 | 0.90 | 0.62 | 0.78 | 0.84 | 0.68 |
| FASL-Seg(Ours) | **0.77** | 0.87 | 0.89 | 0.65 | 0.74 | 0.72 | 0.34 | 0.71 | 0.91 | 0.91 | 0.85 | 0.79 | 0.85 | **0.77** |

We further assess our model's predictions using False Positive Rate (FPR), which is defined as $\frac{FalsePositive+\epsilon}{FalsePositive+TrueNegative+\epsilon}$, where $\epsilon$ is set to 1E-7. The results for each dataset against the SegFormer-b5 model are shown in Table 5. The results show that FASL-Seg has improved FPR results over SegFormer at 1.1% in combined Parts and Anatomy classes, 1.04% for anatomy only and 1% in Tools segmentation. This is attributed to the improved holistic mask generation due to adaptive processing of thin tools as well as larger tools and anatomy through the distinct processing streams.

## 6 Ablation Study

To analyze the effectiveness of the proposed architecture, a thorough ablation study was conducted on the streams' components. First, we analyze the use of attention in the LLFPs and HLFPs, to assess the added value from the multi-head self attention on the feature representations extracted from each encoder output. The investigated configurations and corresponding model performances are presented in Table 6.

The results reveal that applying attention on all the hidden state projections does not necessarily improve the model segmentation ability. Applying attention on HLFP2, which projects the smallest feature map size, shows a drop in the performance compared to no attention applied. This is supported by the knowledge that later encoder blocks encode high-level features; applying attention may result in loss of crucial semantic understanding of the surgical scene. Contrastively, earlier encoder feature maps have more fine-grained knowledge of the image content. Thus, it is observed that attention applied to LLFP1 and LLFP2 improved the mean IoU and Dice compared to no attention. The next ablation study was conducted on the number of attention heads to use in the LLFPs. One, two, and four heads were investigated. The results are presented in Table 7. Initially, the overall performances reveals two head attention performs better than the one or four heads. However, per-class results reveal that some classes were captured better with four-head attention than with two-head and vice versa. An additional experiment was conducted with a combination of two-head attention in LLFP1 and four-head attention in LLFP2, which resulted in the best performance.

An ablation study was conducted on the use of Convolution Transpose (ConvTrans) layers against regular Bilinear interpolation. The ConvTrans were followed by batch normalization and ReLU matching the Conv Blocks used in the rest of the architecture. Surprisingly, using ConvTrans blocks did not provide additional insights on the feature maps, and lead to losing important insights when the features were enlarged. Instead, the use of interpolation was found to improve the model output.

A comparison between the model complexity of FASL-Seg against several SOTA methods is presented in Table 9. Despite some SOTA architectures having more parameters or FLOPS, FASL-Seg was able to outperform them in the three benchmarks. Furthermore, FASL-Seg has lower parameters than SegFormer even with the additional components. The inference speed of our model on the A6000 GPU with 48GB of RAM was 2.14 frames per second, with peak GPU memory at 0.92GB. Thus, the current state of FASL-Seg is more suitable to run post-operative analysis of surgical videos.

**Table 3.** Results for EndoVis18 Tool Segmentation. Per-Class metrics is presented in the form mIoU[Dice]. Label Key: BF: Bipolar Forceps, PF: Prograsp Forceps, LND: Large Needle Driver, SI: Suction Instrument, CA: Clip Applier, MCS: Monopolar Curved Scissors, UP: Ultrasound Probe

| Model | mIoU | Dice | BF | PF | LND | SI | CA | MCS | UP | Avg |
|---|---|---|---|---|---|---|---|---|---|---|
| UNet | 0.64 | 0.66 | 0.66[0.74] | 0.18[0.19] | 0.48[0.50] | 0.74[0.75] | 0.8[0.8] | 0.67[0.72] | 0.62[0.62] | 0.59[0.62] |
| SegFormer | 0.71 | 0.72 | 0.11[0.11] | **0.87[0.87]** | 0.82[0.82] | 0.85[0.85] | **0.95[0.95]** | 0.27[0.27] | **0.97[0.97]** | 0.69[0.69] |
| TraSeTR [25] | - | - | 0.76 | 0.53 | 0.47 | 0.41 | 0.14 | 0.86 | 0.18 | 0.48 |
| S3Net [6] | 0.74 | - | 0.77 | 0.51 | 0.20 | 0.51 | 0.0 | 0.92 | 0.07 | 0.48 |
| MSLRGR [19] | - | - | 0.70 | 0.44 | 0.0 | 0.35 | 0.04 | 0.87 | 0.12 | 0.36 |
| MATIS [5] | 0.84 | - | 0.82 | 0.47 | 0.66 | 0.69 | 0.00 | 0.92 | 0.21 | 0.54 |
| ViTxCNN [23] | 0.85 | 0.83 | **0.86** | 0.68 | 0.73 | **0.89** | 0.06 | 0.91 | 0.22 | 0.64 |
| FASL-Seg(Ours) | **0.86** | **0.87** | 0.79[**0.85**] | 0.70[0.71] | **0.84[0.84]** | 0.80[0.81] | 0.95[0.95] | **0.92[0.95]** | 0.88[0.88] | **0.84[0.85]** |

**Table 4.** Results for EndoVis17 Tool Segmentation. Per-Class metrics is presented in the form mIoU[Dice]. Label Key: BF: Bipolar Forceps, PF: Prograsp Forceps, LND: Large Needle Driver, VS: Vessel Sealer, GR: Grasping Retractor, MCS: Monopolar Curved Scissors, UP: Ultrasound Probe

| Model | mIoU | Dice | BF | PF | LND | VS | GR | MCS | UP | Avg |
|---|---|---|---|---|---|---|---|---|---|---|
| UNet | 0.42 | 0.44 | 0.17[0.19] | 0.16[0.18] | 0.26[0.28] | 0.39[0.39] | 0.52[0.52] | 0.63[0.66] | 0.33[0.33] | 0.35[0.36] |
| SegFormer | 0.67 | 0.68 | 0.46[0.48] | 0.44[0.46] | 0.54[0.56] | 0.63[0.63] | 0.79[0.79] | 0.68[0.7] | 0.87[0.97] | 0.63[0.66] |
| TraSeTR [25] | 0.65 | - | 0.45 | 0.57 | 0.56 | 0.39 | 0.11 | 0.31 | 0.18 | 0.37 |
| S3Net [6] | 0.72 | - | **0.75** | 0.54 | 0.62 | 0.36 | 0.27 | 0.43 | 0.28 | 0.47 |
| MATIS [5] | 0.71 | - | 0.69 | 0.52 | 0.52 | 0.32 | 0.19 | 0.24 | 0.25 | 0.65 |
| ViTxCNN [23] | 0.69 | - | 0.66 | **0.68** | **0.71** | 0.43 | 0.13 | 0.40 | 0.29 | 0.47 |
| FASL-Seg(Ours) | **0.73** | **0.74** | 0.62[0.64] | 0.54[0.55] | 0.63[0.64] | **0.66[0.66]** | **0.81[0.81]** | **0.74[0.76]** | **0.89[0.89]** | **0.70[0.71]** |

**Table 5.** False Positive Rate (FPR) results for the EndoVis18 Parts and Anatomy and Tool Segmentation datasets for FASL-Seg against SegFormer

| Classes on EndoVis18 | SegFormer-b5 (FPR) | FASL-Seg (FPR) |
|---|---|---|
| Parts and Anatomy Segmentation | 0.078 | **0.067** |
| Only Parts | 0.003 | 0.003 |
| Only Anatomy | 0.0704 | **0.06** |
| Tools Segmentation | 0.005 | **0.004** |

**Table 6.** Results for Ablation on Attention Utilization in Feature Processing Streams

| Model | LLFP1 | LLFP2 | HLFP1 | HLFP2 | mIoU | Dice |
|---|---|---|---|---|---|---|
| Model-1 | | | | | 0.6679 | 0.7103 |
| Model-2 | ✓ | | | | 0.6785 | 0.7202 |
| Model-3 | | ✓ | | | 0.6716 | 0.7135 |
| Model-4 | | | | ✓ | 0.6669 | 0.7085 |
| Model-5 | | | ✓ | ✓ | 0.6497 | 0.6909 |
| FASL-Seg | ✓ | ✓ | | | **0.6823** | **0.7236** |

## 7 Broader Impact and Limitations

### 7.1 Broader Impact and Ethical Considerations

While FASL-Seg is designed to enhance surgical situational awareness, its real-time deployment will introduce potential sociotechnical risks. First, highly accurate overlays may foster surgeon over-trust, causing clinicians to accept segmentations uncritically and overlook latent errors. Second, false-positive or false-negative masks could lead to unnecessary cautery, instrument collisions, or missed bleeding points, raising patient-safety and medico-legal concerns. Third, responsibility for adverse outcomes may become ambiguous, complicating liability attribution between the clinical team, hospital, and technology vendor. To mitigate these issues, during future clinical testing, we plan to (i) attach well-calibrated confidence maps to every pixel prediction so that low-certainty regions are visually distinguishable; (ii) maintain a human-in-the-loop paradigm in which overlays are purely advisory—surgeons can toggle or veto them with a foot-switch, and critical actions still require manual confirmation; (iii) log all frames, predictions, and user interactions for post-operative audit trails and algorithm retraining; and (iv) adhere to IEC 62304 and ISO 14971 processes to document hazard analysis, residual risk, and user-training requirements. These safeguards will maximize clinical benefit while minimizing potential harm and ensuring transparent accountability.

### 7.2 Limitations

While FASL-Seg exhibits high performance and introduces new SOTA results, the current work has some limitations. The performance of FASL-Seg was observed on two datasets for a similar operation of nephrectomy, and is yet to be evaluated on additional surgical datasets for other operations. Cross-domain testing has not been conducted to observe the generalizability of this model across various domains. Also, the current model complexity may introduce hardware and time constraints, making it more suitable for post-operative video analysis than real-time deployment. This can be addressed by investigating lightweight backbones and alternative attention mechanisms offering equivalent insights to the model.

## 8 Conclusion

To conclude, this paper proposes a new multiscale segmentation architecture, FASL-Seg, that applies distinct processing mechanisms on varying feature representations extracted using a SegFormer encoder to adapt feature learning based on the resolution of the feature map and the channel-level data. To achieve this, two feature processing streams are proposed: an LLFP stream, which incorporates attention to enhance low-level features and minimize noise, and an HLFP stream to emphasize contextual features. The enhanced feature representation is passed to a shallow decoder to preserve the extracted spatial data, producing the final model output. A comprehensive evaluation was performed using the EndoVis18 and EndoVis17 datasets, for parts, anatomy, and tool segmentation use cases. The model achieved new SOTA performance and exceeded benchmark methods in overall metrics and average per-class metrics, showing FASL-Seg's ability to achieve consistently high performance across various object representations. In future work, we will analyze the architecture components using explainability methods such as GradCAM [20]. Furthermore, we will investigate the use of additional model components to emphasize important feature representations for fusing low-level and high-level features, in addition to lightweight model backbones to achieve similarly high segmentation performance, enabling the use of this architecture in real-time clinical settings.

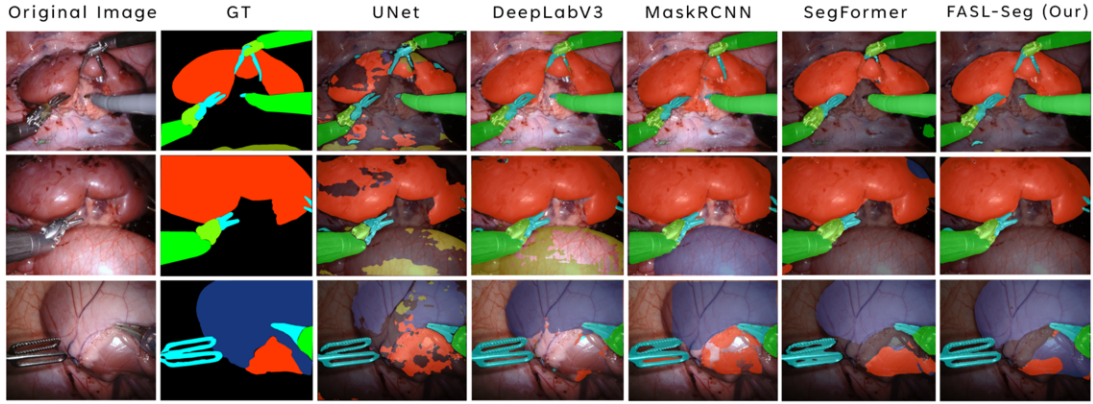

**Figure 5.** Inference Comparison on EndoVis18 Parts and Anatomy Segmentation of FASL-Seg against SOTA

**Table 7.** Ablation on Number of Attention Heads. For per-class, mIoU is presented. Label Key: ISh:Instrument Shaft, IC: Instrument Clasper, IW: Instrument Wrist, SI: Suction Instrument, UP: Ultrasound Probe

| Model | Atten. Heads 1 | 2 | 4 | mIoU | Dice | ISh | IC | IW | Thread | Clamps | Needle | SI | UP |
|---|---|---|---|---|---|---|---|---|---|---|---|---|---|
| Model-6 | ✓ | | | 0.6392 | 0.6803 | 0.82 | 0.522 | 0.585 | 0.22 | 0.74 | 0.905 | 0.754 | 0.772 |
| Model-7 | | ✓ | | 0.6535 | 0.6938 | 0.832 | 0.526 | 0.595 | 0.311 | 0.78 | 0.905 | 0.746 | 0.751 |
| Model-8 | | | ✓ | 0.6441 | 0.6845 | 0.79 | 0.528 | 0.597 | 0.288 | 0.798 | 0.905 | 0.735 | 0.772 |
| FASL-Seg | | ✓ | ✓ | **0.6823** | **0.7236** | **0.847** | 0.526 | **0.649** | **0.483** | **0.799** | **0.905** | **0.811** | **0.802** |

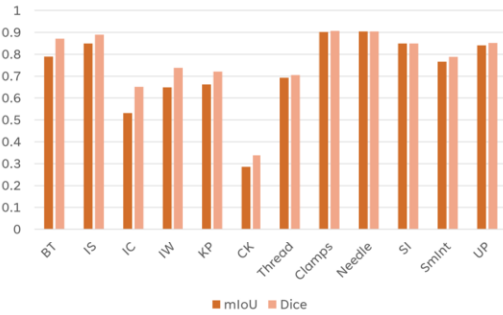

**Figure 6.** Per-Class Metrics for EndoVis18 Parts and Anatomy Segmentation

**Table 8.** Results of Ablation on Upsampling Mechanism used throughout the model architecture

| Model | ConvTrans | Interpolation | mIoU | Dice |
|---|---|---|---|---|
| Model-9 | ✓ | | 0.6823 | 0.7236 |
| FASL-Seg | | ✓ | **0.7222** | **0.7622** |

**Table 9.** Model Complexity of FASL-Seg against SOTA

| Model | Architecture | #Params | GFLOPs |
|---|---|---|---|
| UNet | CNN | 13.39M | 124.44 |
| MaskRCNN | CNN | 44M | 447 |
| DeepLabV3 | CNN | 60.99M | 258.74 |
| TransUNet | Transformer,CNN | 105.3M | 38.6 |
| SegFormer-b5 | Transformer | 84.6M | 110.25 |
| FASL-Seg | Transformer,CNN | 81.99M | 223.42 |

## Acknowledgements

The authors would like to acknowledge the support of the Surgical Research Section at Hamad Medical Corporation for the conduct of this research.

Research reported in this publication was supported by the Qatar Research Development and Innovation Council (QRDI) grant number ARG01-0522-230266. The content reported through this research is solely the responsibility of the authors and does not necessarily represent the official views of Qatar Research Development and Innovation Council.

## References

[1] M. Abdel-Ghani. Muraam-abdel-ghani/fasl-seg: Fasl-seg v1.0.0 – endovis18/17 benchmark implementation, 2025. URL https://doi.org/10.5281/zenodo.16920427.

[2] F. A. Ahmed, M. Yousef, M. A. Ahmed, H. O. Ali, A. Mahboob, H. Ali, Z. Shah, O. Aboumarzouk, A. Al Ansari, and S. Balakrishnan. Deep learning for surgical instrument recognition and segmentation in robotic-assisted surgeries: a systematic review. *Artificial Intelligence Review*, 58(1):1, 2024.

[3] M. Allan, A. Shvets, T. Kurmann, Z. Zhang, R. Duggal, Y.-H. Su, N. Rieke, I. Laina, N. Kalavakonda, S. Bodenstedt, et al. 2017 robotic instrument segmentation challenge. *arXiv preprint arXiv:1902.06426*, 2019.

[4] M. Allan, S. Kondo, S. Bodenstedt, S. Leger, R. Kadkhodamohammadi, I. Luengo, F. Fuentes, E. Flouty, A. Mohammed, M. Pedersen, et al. 2018 robotic scene segmentation challenge. *arXiv preprint arXiv:2001.11190*, 2020.

[5] N. Ayobi, A. Pérez-Rondón, S. Rodríguez, and P. Arbeláez. Matis: Masked-attention transformers for surgical instrument segmentation. In *Proceedings of the IEEE 20th International Symposium on Biomedical Imaging (ISBI)*, 2023. doi: 10.1109/ISBI53787.2023.10230819. URL https://arxiv.org/abs/2303.09514.

[6] B. Baby, D. Thapar, M. E. Chasmai, C. Arora, et al. From forks to forceps: A new framework for instance segmentation of surgical instruments. In *Proceedings of the IEEE/CVF Winter Conference on Applications of Computer Vision (WACV)*, 2023. doi: 10.1109/WACV56688.2023.00613. URL https://doi.org/10.1109/WACV56688.2023.00613.

[7] J. Chen, Y. Lu, Q. Yu, X. Luo, E. Adeli, Y. Wang, L. Lu, A. L. Yuille, and Y. Zhou. Transunet: Transformers make strong encoders for medical image segmentation. *Medical Image Analysis*, 74:102073, 2021. doi: 10.1016/j.media.2021.102073.

[8] L.-C. Chen, G. Papandreou, F. Schroff, and H. Adam. Rethinking

atrous convolution for semantic image segmentation. *arXiv preprint arXiv:1706.05587*, 2017. URL https://arxiv.org/abs/1706.05587.

[9] C. González, L. Bravo-Sánchez, and P. Arbelaez. Isinet: An instance-based approach for surgical instrument segmentation. In A. L. Martel et al., editors, *Medical Image Computing and Computer Assisted Intervention – MICCAI 2020*, volume 12263 of *Lecture Notes in Computer Science*. Springer, Cham, 2020. doi: 10.1007/978-3-030-59716-0_57. URL https://doi.org/10.1007/978-3-030-59716-0_57.

[10] K. He, G. Gkioxari, P. Dollar, and R. Girshick. Mask r-cnn. In *Proceedings of the IEEE International Conference on Computer Vision (ICCV)*, Oct 2017.

[11] Y. Jin, Y. Yu, C. Chen, Z. Zhao, P.-A. Heng, and D. Stoyanov. Exploring intra- and inter-video relation for surgical semantic scene segmentation. *IEEE Transactions on Medical Imaging*, 41(11):2991–3002, 2022. doi: 10.1109/TMI.2022.3177077.

[12] M. Liu, Y. Han, J. Wang, C. Wang, Y. Wang, and E. Meijering. Lskanet: Long strip kernel attention network for robotic surgical scene segmentation. *IEEE Transactions on Medical Imaging*, TMI, 2023. doi: 10.1109/TMI.2023.1234567.

[13] A. Mao, M. Mohri, and Y. Zhong. Cross-entropy loss functions: theoretical analysis and applications. In *Proceedings of the 40th International Conference on Machine Learning*, ICML'23. JMLR.org, 2023.

[14] W. Matasyoh, J. Muthoni, and W. Zhang. Samsurg: Surgical instrument segmentation in robotic surgeries using vision foundation model. *IEEE Access*, 12:12345–12356, 2024. doi: 10.1109/ACCESS.2024.1234567.

[15] Y. Pang, Y. Li, J. Shen, and L. Shao. Towards bridging semantic gap to improve semantic segmentation. In *Proceedings of the IEEE/CVF International Conference on Computer Vision (ICCV)*, pages 4230–4239. IEEE, 2019.

[16] J. N. Paranjape, N. G. Nair, S. Sikder, S. S. Vedula, and V. M. Patel. Adaptivesam: Towards efficient tuning of sam for surgical scene segmentation. *OpenReview*, 2023. URL https://openreview.net/forum?id=aZPRjTqYhv.

[17] O. Ronneberger, P. Fischer, and T. Brox. U-net: Convolutional networks for biomedical image segmentation. In *Medical Image Computing and Computer-Assisted Intervention – MICCAI 2015*, pages 234–241. Springer, 2015. doi: 10.1007/978-3-319-24574-4_28.

[18] S. S. M. Salehi, D. Erdogmus, and A. Gholipour. Tversky loss function for image segmentation using 3d fully convolutional deep networks. In Q. Wang, Y. Shi, H.-I. Suk, and K. Suzuki, editors, *Machine Learning in Medical Imaging*, pages 379–387, Cham, 2017. Springer International Publishing. ISBN 978-3-319-67389-9.

[19] L. Seenivasan, S. Mitheran, M. Islam, and H. Ren. Global-reasoned multi-task learning model for surgical scene understanding. *IEEE Robotics and Automation Letters*, 7(2):314–321, 2022. doi: 10.1109/LRA.2022.3146544.

[20] R. R. Selvaraju, M. Cogswell, A. Das, R. Vedantam, D. Parikh, and D. Batra. Grad-cam: visual explanations from deep networks via gradient-based localization. *International journal of computer vision*, 128:336–359, 2020.

[21] J. M. J. Valanarasu, P. Oza, I. Hacihaliloglu, and V. M. Patel. Medt: A transformer-based medical image segmentation approach. *Medical Image Analysis*, 75:102321, 2021. doi: 10.1016/j.media.2021.102321.

[22] A. Vaswani, N. Shazeer, N. Parmar, J. Uszkoreit, L. Jones, A. N. Gomez, L. Kaiser, and I. Polosukhin. Attention is all you need. In *Advances in Neural Information Processing Systems (NeurIPS)*, pages 5998–6008, 2017. URL https://proceedings.neurips.cc/paper/2017/file/3f5ee243547dee91fbd053c1c4a845aa-Paper.pdf.

[23] M. Wei, M. Shi, and T. Vercauteren. Enhancing surgical instrument segmentation: Integrating vision transformer insights with adapter. *International Journal of Computer Assisted Radiology and Surgery*, 19:1313–1320, 2024. doi: 10.1007/s11548-024-03140-z.

[24] E. Xie, W. Wang, Z. Yu, A. Anandkumar, J. M. Alvarez, and P. Luo. Segformer: Simple and efficient design for semantic segmentation with transformers. In *Advances in Neural Information Processing Systems (NeurIPS)*, 2021. URL https://proceedings.neurips.cc/paper/2021/file/64f1f27bf1b4ec22924fd0acb550c235-Paper.pdf.

[25] Z. Zhao, Y. Jin, and P.-A. Heng. Trasetr: Track-to-segment transformer with contrastive query for instance-level instrument segmentation in robotic surgery. In *Proceedings of the IEEE International Conference on Robotics and Automation (ICRA)*, 2022. doi: 10.48550/arXiv.2202.08453.